\documentclass[aps,prb,twocolumn,groupedaddress,10pt,showpacs,showkeys,preprintnumbers,citeautoscript,floatfix]{revtex4-1}

\usepackage[pdftex]{graphicx}

\usepackage{epstopdf}
\usepackage{dcolumn}
\usepackage{bm}
\usepackage[hidelinks]{hyperref}
\usepackage[mathlines]{lineno}

\usepackage[T1]{fontenc}
\usepackage[latin1]{inputenc}
\usepackage{mathptmx}
\usepackage[protrusion=true,expansion=true]{microtype}	
\usepackage{amsmath,amsfonts,amsthm,amssymb}										
\usepackage{url}
\usepackage{siunitx,booktabs,braket}
\usepackage{listings,slashed}
\usepackage{color}
\usepackage{bigints}
\usepackage{framed}
\usepackage{fancyhdr} 
\usepackage{multirow}
\usepackage[normalem]{ulem} 
\graphicspath{{figure/}}

\def\be{\begin{equation}}
\def\ee{\end{equation}}
\newcommand{\ba}{\begin{eqnarray}}
\newcommand{\ea}{\end{eqnarray}}

\begin{document}


\title{Electron-phonon thermalization in a scalable method for real-time quantum dynamics}
\author{Valerio Rizzi}
\email[Corresponding author. ]{vrizzi01@qub.ac.uk}
\author{Tchavdar N. Todorov}
\author{Jorge J. Kohanoff}
\affiliation{Atomistic Simulation Centre, Queen's University Belfast, Belfast BT7 1NN, Northern Ireland, United Kingdom}
\author{Alfredo A. Correa}
\affiliation{Quantum Simulations Group, Lawrence Livermore National Laboratory, Livermore, California, 94551, USA}

\date{\today}

\begin{abstract}
We present a quantum simulation method that follows the dynamics 
of out-of-equilibrium many-body systems of electrons 
and oscillators in real time. 
Its cost is linear in the number of oscillators 
and it can probe timescales from attoseconds to hundreds of picoseconds.
Contrary to Ehrenfest dynamics, it can thermalize starting 
from a variety of initial conditions, 
including electronic population inversion. 
While an electronic temperature can be defined in terms of a
non-equilibrium entropy, a Fermi-Dirac distribution in general emerges
only after thermalization. 
These results can be used to construct a kinetic model 
of electron-phonon equilibration based on the explicit quantum dynamics.
\end{abstract}


\maketitle

\section{Introduction}
Thermalization between electronic and vibrational degrees of freedom
arises in a range of physical situations
spanning widely different time and length scales.
Examples include Joule heating and dissipation in solid state and molecular physics  
\cite{Horsfield2006,Galperin2006}, equilibration 
of warm dense matter generated 
by laser pulses \cite{Fann1992,Laboratories1992,Ogitsu2012}
and radiation cascades \cite{Duffy2009a} .
The interest in coupled dynamics of 
out-of-equilibrium electrons with vibrations
occurs in several fields, 
including transport in molecular junctions 
\cite{Hartle2011,Lu2012a} and photoelectron spectroscopy \cite{Avigo2013},
and has triggered the development 
of new experimental techniques \cite{Lewis2015}.
Meanwhile, real-time atomistic simulations
venture more and more often into non-equilibrium problems 
where accounting for electron-phonon thermalization is
crucial\footnote{Kogoj {\em et al.} arXiv:1509.08431, which
appeared on arXiv while this paper was under consideration at Physical Review.}.
A choice of methods can capture the interaction
between electrons and vibrations,
from the phenomenological 
Boltzmann equation in extended systems \cite{Ashcroft}
to its counter-part at the nanoscale, non-equilibrium Green's 
functions (NEGF) \cite{Frederiksen2007}.

Nevertheless, the problem of thermal equilibration 
between interacting degrees of freedom (DOF)
is particularly difficult to tackle 
from the simulation point of view.
For purely classical systems simulated via Molecular Dynamics, 
anharmonicities in the potential lead 
- not without technical problems - to thermalization
and energy equipartition \cite{Frenkel}. 
In harmonic or weakly anharmonic systems,
equilibration does not happen spontaneously:
it requires the introduction of external thermostats.
The situation is even more complicated for quantum interacting systems
and it becomes especially critical in mixed quantum-classical approaches.
A widely used approach is the macroscopic two-temperature model
where nuclear and electronic motion is represented in terms
of temperature fields coupled via appropriate diffusion equations
\cite{Anisimov1975,Flynn1988}. This together with 
the introduction of Langevin thermostats \cite{Caro1989}
has proved successful in interpreting 
measured quantities \cite{Finnis1991,Pronnecke1991}.
This approach remains of active interest and, in recent years,
it has evolved into more elaborate methodologies where the nuclear
motion is taken into account via classical molecular dynamics simulations
while electrons are treated at increasing levels of sophistication 
\cite{Duffy2007,Race2010,Mason2015,Cho2011,Karim2014,Zarkadoula2014a}.

The simplest approach to non-adiabatic electron-nuclear atomistic 
simulation is Ehrenfest dynamics (ED) \cite{Horsfield2006} 
in which classical nuclei interact with the mean electron density.
ED is tractable and simple but 
it fails to describe the spontaneous decay of electronic excitations into phonons
because of the lack of microscopic detail in the electronic density 
and resultant loss of electron-nuclear correlation \cite{Horsfield2004}.
Vibrational DOF spontaneously
cool down at the expense of increasing the electronic energy, 
violating the second law of thermodynamics.
What is missing in ED are the collisions 
that drive the probability distribution function towards equilibrium.
The approach to equilibrium can be reinstated via Boltzmann's kinetic theory,
i.e. through phenomenological relaxation dynamics. However
to recover this in microscopic dynamics for a closed system
requires thermostating techniques; for quantum DOF this introduces
an additional layer of complexity.

Correlated electron-ion dynamics (CEID) \cite{Horsfield2004,Horsfield2005}
is a method that was developed to go beyond ED.
It starts from the bare electron-nuclear Hamiltonian and solves it approximately
by a perturbative expansion in powers of nuclear fluctuations about the mean trajectory.
However it scales between quadratically and cubically with the number
of nuclear DOF, 
becoming prohibitive beyond a few 
DOF, along with difficulties
in the choice of closure strategy for the 
hierachy of perturbative equations of motion. 
The computational bottleneck persists in 
alternative expansion strategies for the 
electron-nuclear problem\cite{Stella2007}.

Today there is a new impetus in the study of mesoscale systems, 
as their technological applications and simulation capability meet \cite{Wang2015}. 
These systems mark a difficult middle ground between bulk and
the atomic scale.
There is a serious need for a methodology that includes the mechanisms
of thermal equilibration between electron and phonon DOF,
and at the same time is amenable to computer simulation with present day resources \cite{Cahill2014a}.
This need for an efficient approach to 
the dynamics of thermalization at the mesoscale, 
has motivated us to develop a microscopic 
method for coupled real-time quantum electron-phonon dynamics. 
We refer to it as Effective CEID (ECEID).

ECEID advances beyond CEID in terms of 
conceptual and computational tractability 
by exploiting a different starting point:
a system of electrons and harmonic vibrations, 
coupled by an interaction linear in the generalized displacements. 
This more specialized scenario maintains applicability to the
large family of problems involving harmonic nuclear motion,
while offering important advantages.
This Hamiltonian starts from the 
Born-Oppenheimer level of description, with the role
of the coupling being to generate the non-adiabatic corrections.
By contrast, the old CEID method above had the dual challenge 
of first generating the Born-Oppenheimer behaviour (starting
from the bare full Hamiltonian), and then also going beyond.
Furthermore ECEID employs a non-perturbative closure strategy, 
which enables the coupled electron-phonon dynamics 
to be formulated in terms of a set of variables and equations 
of motion that scale
linearly with the number of vibrational DOF. 
This opens the possibility of tackling problems previously out of reach: 
in test runs we have been able to simulate up to
600 electrons interacting with 100 vibrational DOF
on the picosecond time-scale, on an ordinary workstation.
The next section introduces the method, followed by examples,
and critical comparisons with ED and with a kinetic model in Section III. Section IV
gives a summary and concluding remarks.

\section{The ECEID Method}
\noindent
To describe the ECEID method in its most general form, we
start from the Hamiltonian 
\be
\hat{H} = \underbrace{\hat{H}_{\mathrm{e}} + \sum_{\nu=1}^{N_{\mathrm{o}}} \Big( 
\frac{\hat{P}_\nu^2}{2 M_\nu} + \frac{1}{2} K_\nu \hat{X}^2_\nu \Big)}_{\hat{H}_0} 
- \sum_{\nu=1}^{N_{\mathrm{o}}} \hat{F}_\nu \hat{X}_\nu .
\label{eq:hamone}
\ee
Here $\hat{H}_{\mathrm{e}}$ is a general interacting or non-interacting
many-electron Hamiltonian in the absence of vibrations. 
$\hat{X}_\nu$ and $\hat{P}_\nu$
are displacement and canonical momentum operators for oscillator $\nu$, with
mass $M_\nu$ and spring constant $K_\nu$, coupled linearly to
the electrons via the electronic operator $\hat{F}_\nu$.
$N_{\rm o}$ is the number of blue harmonic vibrational DOF.
Any harmonic Hamiltonian in the vibrational DOF can
be brought into this form through a change of generalized coordinates.

The electronic density matrix (DM)
$\hat{\rho}_\mathrm{e}(t) = \mathrm{Tr}_\mathrm{o}(\hat{\rho}(t))$
is obtained from the full electron-phonon DM $\hat{\rho}(t)$ 
by tracing over the oscillator degrees of freedom and
obeys the effective Liouville equation \cite{Horsfield2004}
\be
\label{eq:rhoeom1}
\dot{\hat{\rho}}_\mathrm{e}(t) = \frac{1}{\mathrm{i} \hbar} \
[ \hat{H}_\mathrm{e} , \hat{\rho}_\mathrm{e}(t) ]  
- \frac{1}{\mathrm{i} \hbar} \ \sum_{\nu=1}^{N_{\rm o}} \
[ \hat{F}_\nu , \hat{\mu}_\nu(t) ]
\ee
where $\hat{\mu}_\nu (t) = \mathrm{Tr}_\mathrm{o}(\hat{X}_\nu\hat{\rho} (t))$.
The full DM can be written exactly as
\begin{eqnarray}
\label{eq:rhofull}
\hat{\rho}(t) & = & \mathrm{e}^{- \frac{\mathrm{i}}{\hbar} \hat{H}_0 t}  \hat{\rho}(0) \mathrm{e}^{ \frac{\mathrm{i}}{\hbar} \hat{H}_0 t}  
-  \frac{1}{\mathrm{i} \hbar} \ \sum_{\nu=1}^{N_{\rm o}}  \ \int_0^t \ 
\mathrm{e}^{\frac{\mathrm{i}}{\hbar} \hat{H}_0 (\tau-t)}\times  \nonumber\\
&& \times [ \hat{F}_\nu \hat{X}_\nu  , \hat{\rho}(\tau) ]  \ 
\mathrm{e}^{-\frac{\mathrm{i}}{\hbar} \hat{H}_0 (\tau-t)} \mathrm{d}\tau.
\end{eqnarray}

We require equations of motion (EOM) for
$\hat{\rho}_\mathrm{e}(t)$ and the mean oscillator occupations
$N_\nu(t) = \mathrm{Tr}(\hat{N}_\nu \hat{\rho}(t))$, 
where $\hat{N}_\nu = \hat{a}^{\dagger}_\nu \hat{a}_\nu$ 
with $\hat{a}^{\dagger}_\nu$ ($\hat{a}_\nu$) the creation 
(annihilation) operator for oscillator $\nu$.
To close the equations, 
we place (\ref{eq:rhofull}) in the 
definition of $\hat{\mu}_\nu(t)$ above 
and make three approximations. 
First, in $\hat{\mu}_\nu(t)$ - but not earlier - we put
$\hat{\rho}(\tau) \approx \hat{\rho}_\mathrm{e}(\tau) 
\hat{\rho}_\mathrm{o} (\tau)$.
This retains electron-phonon correlation exactly 
to lowest order in the coupling
$\hat{ F}_\nu$, and approximately to higher order,
in analogy to the self-consistent Born approximation \cite{Horsfield2006}. 
Second, after taking oscillator traces, we retain only terms diagonal in $\nu$,
suppressing electron-mediated phonon-phonon correlation.
Third, we neglect terms of the form 
$\braket{\hat{a}_\nu \hat{a}_{\nu}}$, $\braket{\hat{a}^\dagger_\nu 
\hat{a}^\dagger_{\nu} }$, retaining
only single-phonon processes and excluding anharmonicity.
From this point, $\nu$ is omitted for simplicity of notation.

These approximations correspond to the low electron-phonon coupling limit and yield
\begin{eqnarray}
\label{eq:mupar}
\hat{\mu}(t) & = & \frac{\mathrm{i}}{M \omega} \ \int_0^t \ 
( N(\tau) + \tfrac{1}{2}  ) \ \mathrm{e}^{\frac{\mathrm{i}}{\hbar} \hat{H}_\mathrm{e} (\tau-t)}  \times \nonumber\\
&& \times [ \hat{F} ,  \hat{\rho}_\mathrm{e}(\tau) ] \ 
\mathrm{e}^{-\frac{\mathrm{i}}{\hbar} \hat{H}_\mathrm{e} (\tau-t)} \ \cos\omega(\tau-t)\ \mathrm{d}\tau
\nonumber\\
& - & \frac{1}{2 M \omega} \ \int_0^t  \ \mathrm{e}^{\frac{\mathrm{i}}{\hbar} \hat{H}_\mathrm{e} (\tau-t)} \times \nonumber \\
&& \times \{ \hat{F}  ,  \hat{\rho}_\mathrm{e}(\tau) \} \ 
\mathrm{e}^{-\frac{\mathrm{i}}{\hbar} \hat{H}_\mathrm{e} (\tau-t)} \ \sin\omega(\tau-t)\ \mathrm d\tau,
\end{eqnarray}
where $\omega = \sqrt{K/M}$ is the oscillator angular frequency.
The full derivation of eq. (\ref{eq:mupar}) is given in Appendix \ref{sec:dermu}.

We calculate $\hat{\mu}(t)$ as follows.
We introduce four auxiliary electronic operators 
$(\hat C_\mathrm c, \hat A_\mathrm c, \hat C_\mathrm s, \hat A_\mathrm s)$ per oscillator, 
defined by
\begin{eqnarray}
\hat{C}_{\mathrm{c}}(t) & = &
\int_0^t  \
( N(\tau) + \tfrac{1}{2}  ) \  
\mathrm{e}^{\frac{\mathrm{i}}{\hbar} \hat{H}_{\mathrm{e}} (\tau-t)}\times \nonumber\\
&& \times [ \hat{F} , \hat{\rho}_\mathrm{e}(\tau) ]
\mathrm{e}^{-\frac{\mathrm{i}}{\hbar} \hat{H}_{\mathrm{e}} (\tau-t)} \ \cos \omega(\tau-t)\ \mathrm d\tau 
\label{eq:Cc}
\end{eqnarray}
\begin{equation}
\hat{A}_{\mathrm{c}}(t) =
\frac{1}{2}\int_0^t \mathrm{e}^{\frac{\mathrm{i}}{\hbar} \hat{H}_{\mathrm{e}} (\tau-t)}
\{\hat{F},\hat{\rho}_\mathrm{e}(\tau)\}\mathrm{e}^{-\frac{\mathrm{i}}{\hbar} \hat{H}_{\mathrm{e}} (\tau-t)}\cos \omega(\tau-t)
\ \mathrm d\tau
\end{equation}
with $\hat{C}_{\mathrm{s}}$ and $\hat{A}_{\mathrm{s}}$
obtained by replacing cosine with sine above. 
They obey the EOM
\begin{eqnarray}
\label{eq:dotcc}											
\dot{\hat{C}}_\mathrm{c} (t) &=& - 
\frac{\mathrm{i}}{\hbar} \ [ \hat{H}_{\mathrm{e}},\hat{C}_\mathrm{c}(t) ] 
+ \omega \hat{C}_\mathrm{s}(t) 
+ (N(t) + \tfrac{1}{2}) [\hat{F},\hat{\rho}_\mathrm{e}(t) ] \\
\dot{\hat{C}}_\mathrm{s} (t) &=& -\frac{\mathrm{i}}{\hbar} \ 
[ \hat{H}_{\mathrm{e}} , \hat{C}_\mathrm{s}(t) ] - \omega \hat{C}_\mathrm{c}(t) \\
\label{eq:dotac}	
\dot{\hat{A}}_\mathrm{c} (t) &=& 
- \frac{\mathrm{i}}{\hbar} \
[ \hat{H}_{\mathrm{e}}, \hat{A}_\mathrm{c}(t) ] 
+ \omega \hat{A}_\mathrm{s}(t)
+ \frac{1}{2} \ \{\hat{F}, \hat{\rho}_\mathrm{e}(t) \} \\
\label{eq:dotas}	
\dot{\hat{A}}_\mathrm{s} (t) &=& -\frac{\mathrm{i}}{\hbar} \
[ \hat{H}_{\mathrm{e}}, \hat{A}_\mathrm{s}(t) ] - \omega \hat{A}_\mathrm{c}(t)
\end{eqnarray}
and, in terms of these quantities, (\ref{eq:mupar}) becomes
\be
\hat{\mu}(t) = \frac{1}{M \omega} ( \mathrm{i} \ \hat{C}_\mathrm{c}(t) - \hat{A}_\mathrm{s}(t) ).
\label{eq:mueom1}
\ee
Analogous steps lead to
\be
\dot{N}(t) = \frac{1}{M \hbar \omega} \
\Big( \mathrm{i} \, \mathrm{Tr}_\mathrm{e}(\hat{F} \hat{C}_\mathrm{s}(t)) 
+ \mathrm{Tr}_\mathrm{e}(\hat{F} \hat{A}_\mathrm{c}(t)) \Big),
\label{eq:Neom1}
\ee
giving an EOM for the oscillator occupation numbers.
This closes the system of EOM.
Energy conservation is discussed in Appendix B.

\subsection{Comparison with Ehrenfest dynamics}
The terms involving 
$[ \hat{F} , \hat{\rho}_{\rm e}(t)]$
are related to the electronic friction 
(an effective dissipative force due to electron-hole excitations by the oscillator),
while those with $\{ \hat{F} , \hat{\rho}_{\rm e}(t)\}$ cause
electronic noise and spontaneous phonon emission \cite{Mozyrsky2002,Todorov2014}.

To see this, consider the above problem
within Ehrenfest dynamics: electrons interacting with a classical oscillator,
with phase $\phi$, slowly varying amplitude $A$, displacement
$X(t) = A \sin (\omega t- \phi)$, and velocity 
$V(t) = \dot{X} (t)$. 
Next, average over $\phi$, to sample
different initial conditions.
The counterpart of the earlier approximations reads
$\langle X(t) X(\tau) \hat{\rho}_{\rm e}(\tau,\phi)\rangle_\phi \approx
\langle X(t) X(\tau)\rangle_\phi \hat{\rho}_{\rm e}(\tau)$,
together with suppression of oscillator position-momentum correlations. 
This produces (\ref{eq:mueom1}) \emph{without} the second term, and with $N$
given by $(N + 1/2)\hbar\omega = M \omega^2 A^2 / 2$. 
The phase-averaged power into the Ehrenfest oscillator, 
$\langle V(t) F(t) \rangle_\phi$ with
$F(t) = \mathrm{Tr}_{\mathrm{e}}(\hat{F} \hat{\rho}_{\rm e}(t,\phi))$,
becomes (\ref{eq:Neom1}) without the second term. 
Finally, the remaining first term in (\ref{eq:Neom1}) is the same as the 
mean rate of work by the electronic friction force due to the symmetric part 
of the velocity-dependent force kernel in equation (16) in \cite{Todorov2014}. 
Thus the ECEID EOM with the anticommutator in (\ref{eq:dotac}) suppressed describe ED
(with oscillator phase averaged out), physically dominated by electron-hole 
excitations and electronic friction.

The second term in (\ref{eq:Neom1}) corresponds instead to the 
power delivered to the oscillators by the effective electronic-noise force 
described by line 1 of equation (56) in \cite{Todorov2014}:
the key correction beyond the mean-field ED. The competition between 
the two terms in (\ref{eq:Neom1}) enables thermodynamic electron-phonon
equilibration \cite{Todorov2014}, which is
thus built into the ECEID method.

\subsection{From many-electron to one-electron equations of motion} 

The EOM above are still many-electron equations. 
To be able to apply the method to systems with large numbers
of electrons, as a practical necessity we express the EOM
in one-electron form. We do this by tracing out all but one electron through
$N_{\mathrm{e}} \mathrm{Tr}_\mathrm{e,2,\dots,N_{\mathrm{e}}}$, where $N_{\mathrm{e}}$
is the number of electrons.
If $\hat{H}_{\mathrm{e}}$ and $\hat{F}$ are one-body operators,
all operators in the EOM can now be replaced by their one-electron counterparts, 
except for the anticommutator term in eq. (\ref{eq:dotac}) $\{\hat{F} , \hat{\rho}_\mathrm{e} \}$.
Following \cite{Horsfield2004}, it transforms into 
\be
\label{eq:antiFDM}
\{\hat{F}^{(1)} (1), \hat{\rho}_\mathrm{e}^{(1)} (1) \} + 2 \mathrm{Tr}_{\mathrm{e,2}} \Big( \hat{F}^{(1)} (2) \hat{\rho}_\mathrm{e}^{(2)} (1,2) \Big)
\ee
where superscripts $^{(1)}$ and $^{(2)}$ denote respectively one- and  two-electron operators.
The simplest decoupling for the two-particle DM is
\be
\hat{\rho}_\mathrm{e}^{(2)}(12,1'2') = \hat{\rho}_\mathrm{e}^{(1)}(11')\hat{\rho}_\mathrm{e}^{(1)}(22')
- \hat{\rho}_\mathrm{e}^{(1)}(12')\hat{\rho}_\mathrm{e}^{(1)}(21') ,
\label{2edm}
\ee
which is valid for independent electrons. Using this in (\ref{eq:antiFDM}), we obtain
$\{\hat{F} , \hat{\rho}_\mathrm{e}(t) \} 
- 2\hat{\rho}_\mathrm{e}(t) \hat{F} \hat{\rho}_\mathrm{e}(t)$,
where now $\hat{\rho}_\mathrm{e}(t)$ is the one-electron DM,
and all other operators are also one-electron operators
\footnote{Here we ignore the additional
term $\hat{\rho}_\mathrm{e}(t)\,\mathrm{Tr}_{\mathrm{e}} ( \hat{F} \hat{\rho}_\mathrm{e}(t) )$.
It corresponds to the so-called ``Hartree'' diagram in NEGF 
treatments of electron-phonon interactions \cite{Frederiksen2007}, 
and is related to motion of the oscillator centroid, a mean-field property. 
This term involves the mean force 
$\mathrm{Tr}_{\mathrm{e}} ( \hat{F} \hat{\rho}_\mathrm{e}(t) )$ 
on a given degree of freedom,
which in the present examples is orders of magnitude
less than a typical interatomic bond force.}.
The accuracy of (\ref{2edm}) reduces with increased
electron-phonon coupling; corrections are discussed in \cite{Horsfield2005}.

Screening can be included in a one-electron mean-field picture 
within a Hartree-Fock scheme following \cite{Horsfield2005}, 
or in a time-dependent density-functional framework \cite{Burke2005}. 

To simulate a finite system, we must account for the 
level-broadening and decoherence introduced by the environment.
We replace
$[ \hat{H}_{\mathrm{e}} , \hat{Q} ]$ in 
(\ref{eq:dotcc}-\ref{eq:dotas}) by
$\hat{H}_\Gamma \hat{Q} - \hat{Q} \hat{H}_\Gamma^{\dagger}$ 
where $\hat{Q} = (\hat C_\mathrm c, \hat A_\mathrm c, \hat C_\mathrm s, 
\hat A_\mathrm s)$, 
$\hat{H}_\Gamma = \hat{H}_{\mathrm{e}} - 
\mathrm{i} \Gamma \hat{I}_{\mathrm{leads}}$,  
and  $\hat{I}_{\mathrm{leads}}$ is the identity 
operator in the leads with $\Gamma$ a small positive quantity. 
The total energy
$E = E_\mathrm{e} + E_\mathrm{o} + E_\mathrm{c}$, where 
$E_\mathrm{e} = \mathrm{Tr}_{\rm e}(\hat{H}_{\rm e} \hat{\rho}_{\rm e}(t))$,
$E_\mathrm{o} = \sum_\nu \hbar \omega_\nu ( N_\nu (t) + 1/2 )$ and
$E_\mathrm{c} = - \sum_{\nu}\mathrm{Tr}_{\rm e}(\hat{F}_\nu \hat{\mu}_\nu (t))$,
is identically conserved, provided the damping self-energy and the electron-phonon 
coupling $\hat{F}$ lie in different subspaces. 
The derivation of this result is shown in Appendix \ref{sec:consen}.
In our examples, once $\Gamma$ exceeds the 
energy-level spacing in the system, 
transition rates resulting
from ECEID dynamics become independent of $\Gamma$.
The role of $\Gamma$ is to mimic an extended 
(infinitely large) system without the extra cost.

\section{Results}
\noindent
Here we have implemented the ECEID method for the discretized 
electron-phonon Hamiltonian (\ref{eq:htb})
\begin{eqnarray}
\label{eq:htb}
\hat H_\mathrm{\rm e-ph} &=& \overbrace{\sum_{ij} \alpha_{ij} \hat c^\dagger_i 
\hat c_j }^{\hat H_\text e} 
- \sum_{\nu ij} 
\overbrace{
F_{\nu i j} \hat c^\dagger_i \hat c_j \ 
\frac{\hat a^\dagger_\nu + \hat a_\nu}
{\sqrt{2 M_\nu \omega_\nu / \hbar}}
}^{\hat F_\nu \hat X_\nu} \\ \nonumber
&+& \sum_\nu \hbar\omega_\nu \Big(\hat a^\dagger_\nu \hat a_\nu + \frac12 \Big) 
\end{eqnarray}
where $\hat c^\dagger$($\hat c$) are the fermion creation (annihilation) operators.
$\alpha_{ij}$ are onsite energies and hoppings with $\{i,j\}$ running over the atomic sites.  
The electronic DM evolves according to eq. (\ref{eq:rhoeom1}). 
$\hat{\mu}_\nu(t)$  is calculated using eq. (\ref{eq:mueom1}),
which is obtained from the time evolution 
of the auxiliary operators (\ref{eq:dotcc}-\ref{eq:dotas}). 
These enter also in the EOM for the mean oscillator occupation (\ref{eq:Neom1}). 
The number of EOM scales linearly with $N_\mathrm{o}$ and
so does the computational cost.

		\begin{figure}[h!tb]
		\includegraphics[width=\columnwidth]{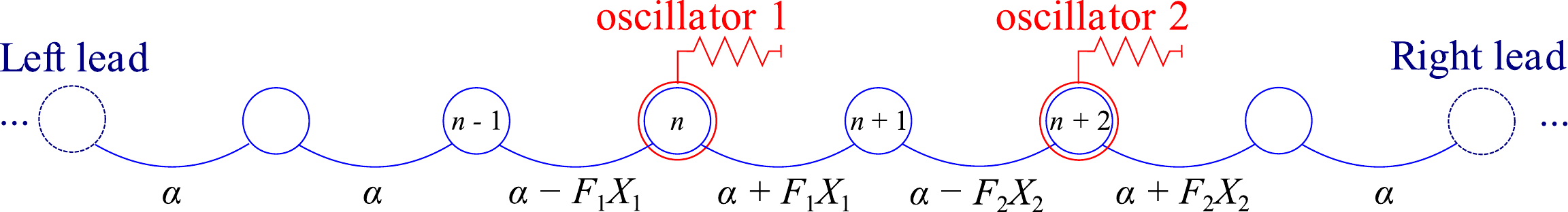}
		\caption{\label{fig:wire}
		Schematic of our model system:
		a nearest-neighbour one-dimensional lattice model of
		an atomic wire divided into a central region between two leads.
		This embeds the sample in an 
		environment, and provides the framework for
		future transport calculations.
	  Each of the 3 regions has 32 sites, with 15 equispaced 
		harmonic oscillators coupled to the central region.
		Oscillator $\nu$ couples to site $n_\nu$ through 
		$\hat F_\nu = F_\nu \Big(\hat c^\dagger_{n_{\nu} + 1} \hat c_{n_{\nu}} 
		+ \hat c^\dagger_{n_{\nu}} \hat c_{n_{\nu} + 1} - \hat c^\dagger_{n_{\nu}} \hat c_{n_{\nu} -1} 
		- \hat c^\dagger_{n_{\nu} - 1} \hat c_{n_{\nu}} \Big)$
		which corresponds to independent atomic motion 
		in a lattice description. The extension from Einstein oscillators to normal modes is straightforward.
		The onsite energies are uniform, the hoppings $\alpha = -1~\mathrm{eV}$
		and $\Gamma = 0.08~\mathrm{eV}$. 
		For all the oscillators $M = 0.5$ a.m.u., 
		$\hbar \omega = 0.2~\mathrm{eV}$ and $F=0.3~\mathrm{eV/\text{\normalfont\AA}}$.
		}
		\end{figure}	

We use these equations to simulate 
non-equilibrium electron-phonon dynamics in the
model in Fig. \ref{fig:wire}: a wire with 
an electronic half-filled band with 96 spin-degenerate non-interacting electrons
coupled to 15 harmonic oscillators.
The integration of the EOM is highly efficient
and parallelizable over the different oscillators.
On a modern 20 processor machine, 
a 10 ps simulation requires about one hour.

To track the evolution of the two subsystems,
we use two temperature-like parameters: $T_{\rm o}^{\mathrm{quant}}$ for the oscillators 
and $T_{\rm e}$ for electrons. If $\overline{N}(t) = \sum_{\nu=1}^{N_{\rm o}} N_{\nu}(t) / N_{\mathrm{o}}$, 
then the oscillator temperature is defined through
$\overline{N}(t) = (\mathrm{e}^{\hbar\omega/k_\text{B} T_{\mathrm{o}}^{\mathrm{quant}}(t)} - 1 )^{-1}$. 
In the Ehrenfest case, this definition breaks down when the energy of the 
classical oscilators goes down to 0 and 
$\overline{N}(t) \rightarrow -1/2$. 
For that case, we employed an alternative semiclassical definition of oscillator 
temperature $k_\text{B} T_{\mathrm{o}}^{\mathrm{class}} = ( \overline{N}(t) + \frac{1}{2} ) \hbar \omega$.
The electronic temperature is taken from $T_{\rm e} = \Delta E_\mathrm{e} / \Delta S_\mathrm{e}$ 
where $\Delta E_\mathrm{e}$ is
the variation over 5 timesteps in electronic energy
and $\Delta S_\mathrm{e}$ is the corresponding variation in electronic 
von Neumann entropy 
$S_\mathrm{e} = -k_{\rm B} \sum_n \big( f_n \log f_n
+ (1-f_n) \log(1-f_n)\big)$, 
where $f_n$ are the diagonal elements of $\hat{\rho}_\mathrm{e}$ 
in the basis of $\hat{H}_{\mathrm{e}}$ eigenstates,
the occupations of the unperturbed electronic energy levels.
$T_\mathrm{e}$ is then inferred from a running average 
of its reciprocal. 
We note that these temperatures are only observables,
not an input into the simulation.

As the system evolves, no macroscopic work is done, but
energy (heat) is exchanged between the electronic and the 
oscillators subsystems.
Having a microsocpic definition of the entropy 
also allows us to give a time-local 
quantification of the rate of heat exchange 
$J_Q = \frac{dS_\text{total}}{dt}/(1/T_\mathrm o - 1/T_\mathrm e)$,
where $dS_\text{total} = dS_\mathrm o + dS_\mathrm e$.
In the weak-coupling limit, where the correlation energy $E_{\mathrm{c}}$ is small, 
the heat current reduces to $J_Q = dE_\mathrm{o}/dt$, 
and on average $dE_\mathrm{o}/dt = - dE_\mathrm{e}/dt$.

\subsection{Thermalization}
\begin{figure}[h!tb]
		\includegraphics[width=\columnwidth]{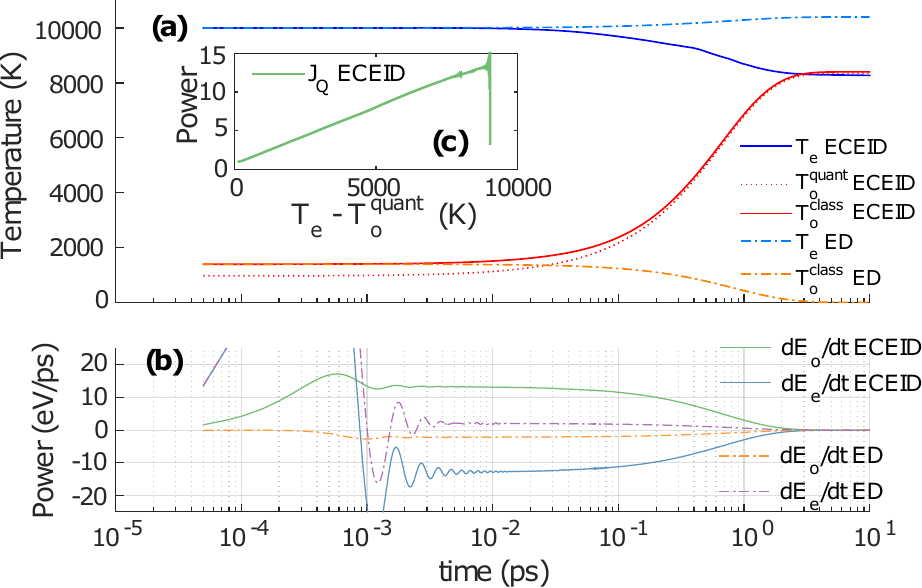}
		\caption{\label{fig:temper} Coupled dynamics of a closed system 
		of electrons and oscillators with the parameters given in the text.
		(a) Time evolution of the electronic and oscillators temperature 
		for ECEID and the phase-averaged ED discussed above.
		(b) Rate of change of electronic and oscillators energies.
		After a transient of $10~\mathrm{fs}$, the systems evolve 
		until eventually an equillibrium state (ECEID) or an unphysical state (ED) is reached.
		(c) For ECEID a clear linear scaling (Fourier law behavior) is observed for heat flow vs. temperature difference (up to a time of 2.5 ps). 
		The noise for high temperature differences is related to the initial transient.}
		\end{figure}

Our first example starts with $T_{\mathrm{e}}=10000$ K 
and $T_{\mathrm{o}}^{\mathrm{class}}=1400$ K. 
This mimics a common situation in laser or irradiation 
experiments in which electrons initially absorb energy faster that ions.
In Fig. \ref{fig:temper} we compare the 
time evolution of the temperature for ED and ECEID.
After a short transient which depends on the details
of the initial state, a long-lived steady state develops 
with a net energy flow from one subsystem to the other.	
In ED, the absence of electronic noise (second term in eq. (\ref{eq:Neom1})) results in a
heat flow going in the wrong direction: 
from the cold oscillators into the hot electrons, until the oscillators 
reach green temperature.
In ECEID, the inclusion of the electronic noise makes the exchange 
of heat physical and the final thermalization possible (Fig.~\ref{fig:temper}(a)).
The heat flow scales linearly with the temperature difference (Fourier's law) (Fig.~\ref{fig:temper}(c)).
In the equilibrium state reached in ECEID, the two final temperatures
agree within 1\%.

\subsection{Population inversion}
\begin{figure}[h!tb]
		\includegraphics[width=\columnwidth]{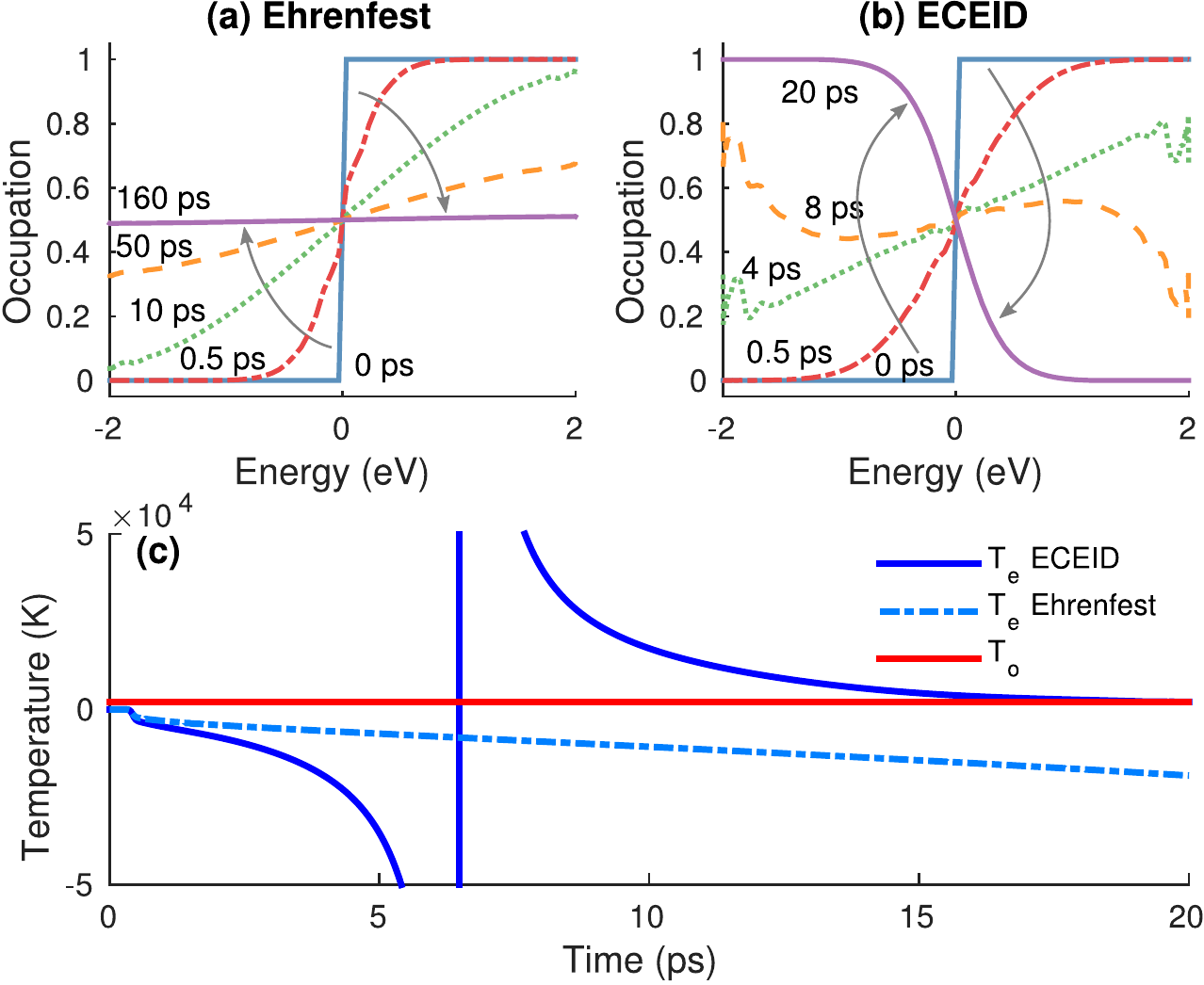}
		\caption{\label{fig:popinv} Population inversion simulation with the oscillators 
		held at constant temperature. 
		We show snapshots of the population of the electronic states in 
		(a) ED at 0 ps, 0.5 ps, 10 ps, 50 ps, 
		160 ps and (b) ECEID at 0 ps, 0.5 ps, 4 ps, 
		8 ps, 20 ps. (The arrows highlight 
		the overall initial-to-final transition in each case.)
	  (c) Temperature evolution during the simulation 
		for ED and ECEID compared with the fixed oscillator temperature.}
		\end{figure}

Next, we test an extremely out-of-equilibrium phenomenon: 
a complete population inversion. 
Initially, the electrons occupy the upper half of the energy states 
in the wire, corresponding to an infinitesimal negative 
electronic temperature. The oscillators are held at $N=0.5$,
or $T_{\mathrm{o}}^{\mathrm{quant}}=2112$ K
throughout. This simulates
coupling to an infinitely efficient
external thermostat, thus 
isolating just the electron dynamics.
Fig. \ref{fig:popinv} shows snaphots 
of the electronic population dynamics and the temperature.
The electrons de-excite in both ECEID and ED.
In ED this happens through negative friction \cite{Lu2011}. 
Comparing Fig. \ref{fig:popinv}(a) and (b)
at 0.5 ps, we see that the de-excitation is faster in ECEID; this is because ECEID
includes also the contribution from spontaneous phonon emission. But the crucial difference
is the final state:
ECEID correctly takes the electrons all the way down to a Fermi-Dirac distribution
corresponding to the oscillator temperature; ED by contrast gets stuck at a distribution
with roughly uniform occupancies \cite{Theilhaber1992}. 
These two ED features have a common origin. 
If electronic occupancies $f(E)$ depend only on energy $E$, 
then a rearrangement of the result for the electronic friction in \cite{Lu2012a} 
gives an integral containing $f'(E)$ as a factor in the integrand. Hence the
opposite signs for the friction, at small negative and small positive 
temperatures. Hence also the unphysical 
``equilibration'' of the electrons at $f'(E) = 0$ in ED,
when the friction vanishes and the main
electron-phonon interaction mechanism present in ED goes to zero.

The role of $\Gamma$ in these simulations is crucial for thermalization 
because it provides a controlled way to embed a finite system, 
that would not equilibrate, into an extended one that does. 
In Fig. \ref{fig:toy} we study the time evolution of a sample 
of electronic states in ECEID for $\Gamma = 0.08$ eV and $\Gamma = 0.8$ eV 
for the same initial population inversion as above.
The results are almost superimposable: 
for $\Gamma$ larger than the average level spacing $\sim 0.04$ eV, 
ECEID is largely independent of $\Gamma$.
We observed that the dynamics of any level $j$ is exactly 
symmetric with that of level $96-j+1$ for all times.

\subsection{Kinetic model}
		\begin{figure}[h!tb]
		\includegraphics[width=\columnwidth]{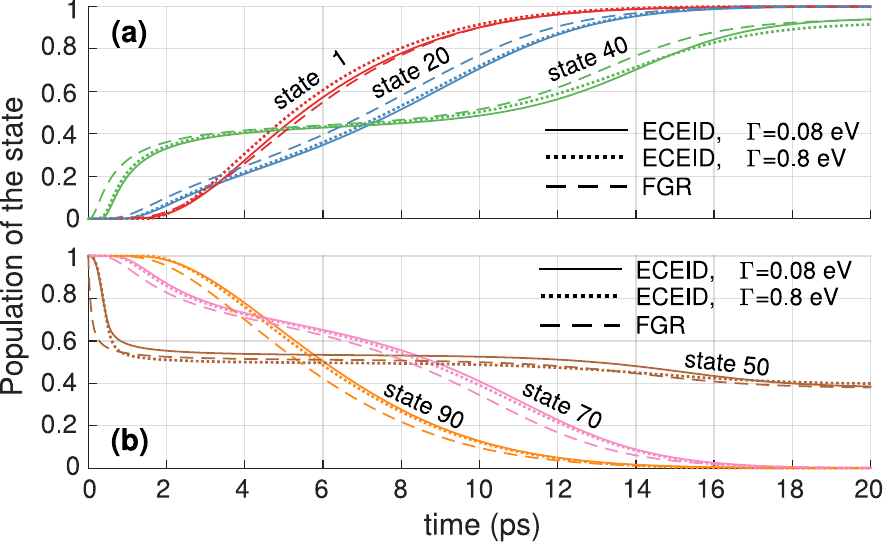}
		\caption{\label{fig:toy} Comparison of the dynamics of electronic states for ECEID 
		with $\Gamma = 0.08$ eV, ECEID with $\Gamma = 0.8$ eV 
		and the kinetic model starting from an inverted population. 
		In (a) we track state 1, state 20 and state 40; 
		in (b) state 90, state 70 and state 50.}
		\end{figure}	
The rich pattern of population evolutions shown in Fig. \ref{fig:toy} 
can be understood with a kinetic model of the 
transitions between electronic levels due to phonon absorption and emission.
The rate equation for the population $f_{j}$ of level $j$ is
\begin{eqnarray}
\dot{f}_j (t) = && \sum_k \frac{1}{\tau_{jk}} ( - N f_j (1 - f_k) \ + (N+1) f_k (1 - f_j) ) \nonumber\\
								&+& \frac{1}{\tau_{kj}} ( N f_k (1 - f_j) \ - (N+1) f_j (1 - f_k) ).
\label{eq:toyrate}											
\end{eqnarray}
The scattering rates $1/\tau_{jk} = (\pi/M \omega) N_{\mathrm{o}} | F_{jk}  |^2  G_{jk}$
are given by the Fermi Golden Rule (FGR).
$|F_{jk}|^2$ can be calculated analytically
by using plane wave states with energies $E_j = 2\alpha\cos{\phi_j}$,
(dimensionless) crystal momentum 
$\phi_j = j \pi/97$, $j=1,\dots,96$ and by averaging over 
the two opposite signs of momentum for the final state. 
$G_{jk} =  \mathrm{e}^{-\left(  (E_k - E_j - \hbar \omega)/\Delta \right)^2 } / ( \sqrt{\pi} \Delta)$
is a Gaussian envelope with a width $\Delta$. It mimics the $\delta$-function that appears 
in the FGR electron-phonon transition rates.
We plug the parameters of the population inversion simulation 
from Fig. \ref{fig:popinv} into the kinetic model with $\Delta = 0.08$ eV 
and in Fig. \ref{fig:toy} we compare it with ECEID simulations,
showing close agreement.

The comparison with the kinetic model illustrates that ECEID, owing to its scalability,
can access time- and size-domains where macroscopic thermodynamic behaviour is beginning to emerge.
In addition, the direct comparison between inherently different descriptions provides a bottom-up 
path to a validation, at the atomistic level, of kinetic models of electron-phonon dynamics, 
without having to resort to the relaxation-time approximation.

The response is fastest for states in the middle of the band, where the step
in the initial population is. The time that these states take to settle into a
long-lived half-occupied steady state - about 0.5 ps - is comparable to
the time needed for the initial temperature response - 
the small initial step-like feature in the 
blue results in Fig. \ref{fig:popinv}(c). (This transient response
in the electron-phonon dynamical simulation is absent in FGR,
because FGR by construction describes mean transition rates
in the long-time limit.)
The results of the kinetic model show little variation over the range $0.04 < 
\Delta < 0.15$ eV or for different shapes of $G_{jk}$. For this choice of
parameters, the kinetic model captures the main
physics of the problem. The combination of the kinetic model and ECEID
provides a direct way to construct rate equations that allow thermodynamic
electron-phonon equilibration on the basis of a real-time quantum
mechanical simulation.

\section{Conclusion}
\noindent
Our method can track the dynamics of interacting
out-of-equilibrium quantum many-body systems of electrons and oscillators
in real time. We have applied it to nanowires from a range of
initial conditions, to demonstrate its ability to describe
thermalization. We show how an entropic
definition of temperature, combined with the microscopic ECEID dynamics,
produces a thermodynamically meaningful description of the
energy exchange between the two sybsystems, and their equilibration. 
A key aspect of the method is the linear scaling with the number
of vibrational DOF. This makes it possible to access large size- and 
time-domains where macroscopic transition dynamics is beginning 
to emerge, and where ECEID provides a basis for suitable kinetic models. 
In contrast with kinetic models, 
ECEID applies also to problems dominated by quantum coherence,
such as electron transport in atomic-scale open systems. An implementation 
of ECEID for current-carrying systems is under development.

\begin{acknowledgments}
We thank Lorenzo Stella and Kieron Burke for helpful discussions.
We express our gratitude to the Leverhulme Trust for funding this research under grant RPG-2012-583.
Work by VR (during a visit hosted by AAC) and by AAC performed under the auspices of the 
U.S. Department of Energy by Lawrence Livermore National Laboratory 
under Contract DE-AC52-07NA27344.
\end{acknowledgments}

\appendix
\section{Derivation of $\hat{\mu} (t)$}
\label{sec:dermu}
For convenience, below we use the notation
\be
\hat{Q}^t = \mathrm{e}^{\frac{\mathrm{i}}{\hbar} \hat{H}_0 t}  \hat{Q} \mathrm{e}^{-\frac{\mathrm{i}}{\hbar} \hat{H}_0 t}
\label{eq:Qinter}
\ee
for a generic operator $\hat{Q}$.
Inserting eq. (\ref{eq:rhofull}) into the definition of $\hat{\mu}_\nu (t)$, we get
\be
\label{eq:muap1}
\hat{\mu}_\nu(t) = - \frac{1}{\mathrm{i} \hbar} \ \mathrm{Tr}_\mathrm{o} \ \Big(   \hat{X}_\nu \sum_{\tilde{\nu}=1}^{N_{\rm o}} 
\int_0^t [ \hat{F}^{\tau-t}_{\tilde{\nu}} \hat{X}^{\tau-t}_{\nu'} , \hat{\rho}^{\tau-t} (\tau) ] \
\mathrm{d}\tau \Big).
\ee
Here we assume for simplicity that the unperturbed 
motion described by $\hat{\rho}^{-t}(0)$ in (\ref{eq:rhofull}) 
does not contribute to motion of the oscillator centroids 
and to $\hat{\mu}(t)$.

We expand the commutator and permute the operators within the oscillator trace in eq. (\ref{eq:muap1}) to obtain
\begin{eqnarray}
\label{eq:mu2}
\hat{\mu}_\nu (t) = - \frac{1}{\mathrm{i} \hbar} \ \mathrm{Tr}_\mathrm{o} \sum_{\nu'=1}^{N_{\rm o}} \Big( && \int_0^t 
 \hat{F}^{\tau-t}_{\nu'}  \hat{X}_{\nu} \hat{X}^{\tau-t}_{\nu'} \hat{\rho}^{\tau-t} (\tau) \  \mathrm{d}\tau \nonumber\\
 - && \int_0^t  \hat{\rho}^{\tau-t} (\tau) \hat{X}^{\tau-t}_{\nu'} \hat{X}_{\nu} \hat{F}^{\tau-t}_{\nu'} \ \mathrm{d}\tau \Big).
\end{eqnarray}
By time-differentiating $\hat{X}^{\tau-t}_{\nu'}$ twice and using the canonical position-momentum commutation relation
$[\hat{X}_{\nu},\hat{P}_{\nu'}] = \mathrm{i} \hbar \delta_{\nu \nu'}$, we can see that
\be
\label{eq:Xddot}
\ddot{\hat{X}}^{\tau - t}_\nu = -\frac{K_\nu}{M_\nu} \hat{X}^{\tau - t}_\nu.
\ee
The solution of (\ref{eq:Xddot}), with the initial conditions $\hat{X}^0_\nu=\hat{X}_\nu$ and $\dot{\hat{X}}^{0}_\nu=\hat{P}_\nu/M_\nu$ 
and with the introduction of the characteristic oscillator frequency $\omega_\nu = \sqrt{K_\nu/M_\nu}$, is
\be
\label{eq:icsti}
\hat{X}^{\tau - t}_\nu = \hat{X}_\nu \cos \omega_\nu (\tau - t) + \frac{\hat{P}_\nu}{M_\nu \omega_\nu} \sin \omega_\nu (\tau - t),
\ee
which can be rewritten in second quantization as 
\be
\label{eq:ics2q}
\hat{X}^{\tau-t}_\nu = \sqrt{\frac{\hbar}{2M_\nu\omega_\nu}} (\hat{a}^\dagger_\nu 
\mathrm{e}^{\mathrm{i} \omega_\nu (\tau-t)}  + \hat{a}_\nu \mathrm{e}^{-\mathrm{i} \omega_\nu (\tau-t)} ).
\ee
Now we apply the decomposition 
\be
\label{eq:AB}
\hat{A} \hat{B} = \frac{1}{2} \{ \hat{A},\hat{B} \} + \frac{1}{2} [ \hat{A},\hat{B} ]
\ee
to both $\hat{X}_{\nu} \hat{X}_{\nu'}^{\tau-t}$ 
and $\hat{X}_{\nu'}^{\tau-t} \hat{X}_{\nu}$ 
in eq. (\ref{eq:mu2}), leading to
\begin{align}
\label{eq:mu4}
\hat{\mu}_{\nu} (t) = - \frac{1}{2 \mathrm{i} \hbar} \ \mathrm{Tr}_\mathrm{o} \sum_{\nu'=1}^{N_{\rm o}}  \Big( &\int_0^t 
[ \hat{F}^{\tau-t}_{\nu'},\hat{\rho}^{\tau-t} (\tau)]  \{ \hat{X}_\nu,\hat{X}^{\tau-t}_{\nu'} \} \  \mathrm{d}\tau \Big) \nonumber\\
- \frac{1}{2 M_\nu \omega_\nu} \ \mathrm{Tr}_\mathrm{o} \ \Big( \int_0^t  \{&\hat{F}^{\tau-t}_{\nu}, \hat{\rho}^{\tau-t} (\tau) \} \sin \omega_{\nu} (\tau - t) \ \mathrm{d}\tau \Big),
\end{align}
where we have used 
\be
\label{eq:anticlosed}
\sum_{\nu'=1}^{N_{\rm o}} [\hat{X}_\nu,\hat{X}_{\nu'}^{\tau - t}] 
= \frac{\mathrm{i} \hbar}{M_\nu \omega_\nu} \sin \omega_\nu (\tau - t) .
\ee
Eq. (\ref{eq:mu4}) is exact.

The approximation $\hat{\rho}(\tau) \approx \hat{\rho}_\mathrm{e}(\tau) 
\hat{\rho}_\mathrm{o} (\tau)$ given in the main text is now made
in eq. (\ref{eq:mu4}), yielding
\begin{align}
\label{eq:mu5}
\hat{\mu}_{\nu} (t) = - \frac{1}{2 \mathrm{i} \hbar} \ \sum_{\nu'=1}^{N_{\rm o}}  &\int_0^t 
[ \hat{F}^{\tau-t}_{\nu'},\hat{\rho}^{\tau-t}_{\mathrm{e}} (\tau)] \mathrm{Tr}_\mathrm{o} \Big( \hat{\rho}_{\mathrm{o}}^{\tau-t} (\tau) \{ \hat{X}_\nu,\hat{X}^{\tau-t}_{\nu'} \} \Big) \ \mathrm{d}\tau  \nonumber\\
- \frac{1}{2 M_\nu \omega_\nu} \  \int_0^t  \{&\hat{F}^{\tau-t}_{\nu}, \hat{\rho}_{\mathrm{e}}^{\tau-t} (\tau) \} \sin \omega_{\nu} (\tau - t) \ \mathrm{d}\tau.
\end{align}
Next we take the oscillator trace, making the remaining approximations, namely
retaining only terms diagonal in $\nu$ and ignoring the double (de)excitations 
$\mathrm{Tr}_\mathrm{o} ( \hat{a}_\nu \hat{a}_{\nu} \hat{\rho}_{\mathrm{o}}^{\tau-t} (\tau) )$, 
$\mathrm{Tr}_\mathrm{o} ( \hat{a}^\dagger_\nu \hat{a}^\dagger_{\nu} 
\hat{\rho}_{\mathrm{o}}^{\tau-t} (\tau))$. This gives
\be
\mathrm{Tr}_\mathrm{o} \sum_{\nu'=1}^{N_{\rm o}} \Big(   \{ \hat{X}_\nu,\hat{X}^{\tau-t}_{\nu'} \} \hat{\rho}_{\mathrm{o}}^{\tau-t} (\tau) \Big) 
\approx \frac{\hbar}{M_\nu \omega_\nu} (2 N_\nu(\tau) + 1) \cos \omega_\nu (\tau - t)
\ee 
Then eq. (\ref{eq:mu5}) becomes
\begin{eqnarray}
\label{eq:mu6}
\hat{\mu}_\nu (t) &=& \frac{\mathrm{i}}{M_\nu \omega_\nu} \ \int_0^t 
 \Big(N_\nu(\tau) + \frac{1}{2}\Big) [\hat{F}^{\tau-t}_\nu,\hat{\rho}_{\mathrm{e}}^{\tau-t} (\tau)] \cos \omega_\nu (\tau - t) \ \mathrm{d}\tau \nonumber\\
&-& \frac{1}{2 M_\nu \omega_\nu} \ \int_0^t  \{\hat{F}^{\tau-t}_\nu,
\hat{\rho}_{\mathrm{e}}^{\tau-t} (\tau) \} \sin \omega_\nu (\tau - t) \ \mathrm{d}\tau.
\end{eqnarray}
This is eq. (\ref{eq:mupar}).

\section{Total energy conservation}
\label{sec:consen}
\noindent
The time-derivative of the total energy of the system is
\be
\dot{E} = \mathrm{Tr}_{\rm e}(\hat{H}_{\rm e} \dot{\hat{\rho}}_{\rm e}(t)) + \sum_\nu \left( \hbar \omega_\nu \dot{N}_\nu (t) - \mathrm{Tr}_{\rm e}(\hat{F}_\nu \dot{\hat{\mu}}_\nu (t)) \right).
\label{eq:edot}
\ee
Plugging eq. (\ref{eq:rhoeom1}) into the first term of eq. (\ref{eq:edot}) and
using eq. (\ref{eq:mueom1}), we get
\be
-\frac{1}{M_\nu \hbar \omega_\nu} \mathrm{Tr}_{\rm e}\left( [\hat{F}_\nu,\hat{C}_c^\nu] \hat{H}_{\rm e} + \mathrm{i} [\hat{F}_\nu,\hat{A}_s^\nu] \hat{H}_{\rm e} \right)
\label{eq:primoedot}
\ee
and, with eq. (\ref{eq:Neom1}), the second term becomes
\be
\frac{1}{M_\nu} \mathrm{Tr}_{\rm e} \left( \mathrm{i} \hat{F}_\nu \hat{C}_s^\nu  + \hat{F}_\nu \hat{A}_c^\nu \right).
\label{eq:secondoedot}
\ee
Using the time derivative of eq. (\ref{eq:mueom1}) together with eq. (\ref{eq:dotcc}), (\ref{eq:dotas}) the third term is
\begin{eqnarray}
\label{eq:terzoedot}
&-&\frac{1}{M_\nu \hbar \omega_\nu} \mathrm{Tr}_{\rm e}\left( \hat{F}_\nu ( \hat{H}_{\Gamma} \hat{C}_c^\nu - \hat{C}_c^\nu \hat{H}_{\Gamma}^\dagger) + \mathrm{i} \hat{F}_\nu 
(\hat{H}_{\Gamma} \hat{A}_s^\nu - \hat{A}_s^\nu \hat{H}_{\Gamma}^\dagger) \right) \nonumber\\
&-&\frac{1}{M_\nu} \mathrm{Tr}_{\rm e} \left( \mathrm{i} \hat{F}_\nu \hat{C}_s^\nu  + \hat{F}_\nu \hat{A}_c^\nu \right).
\end{eqnarray}
So long as $\hat{I}_{\mathrm{leads}} \hat{F}_\nu  = 0$, we can replace
$\hat{H}_{\Gamma}$ in eq. (\ref{eq:terzoedot}) by $\hat{H}_{\rm e}$.
Then summing (\ref{eq:primoedot}), (\ref{eq:secondoedot}) and (\ref{eq:terzoedot}) gives $\dot{E} = 0$.

\bibliography{papers,extra}

\begin{thebibliography}{36}%
\makeatletter
\providecommand \@ifxundefined [1]{%
 \@ifx{#1\undefined}
}%
\providecommand \@ifnum [1]{%
 \ifnum #1\expandafter \@firstoftwo
 \else \expandafter \@secondoftwo
 \fi
}%
\providecommand \@ifx [1]{%
 \ifx #1\expandafter \@firstoftwo
 \else \expandafter \@secondoftwo
 \fi
}%
\providecommand \natexlab [1]{#1}%
\providecommand \enquote  [1]{``#1''}%
\providecommand \bibnamefont  [1]{#1}%
\providecommand \bibfnamefont [1]{#1}%
\providecommand \citenamefont [1]{#1}%
\providecommand \href@noop [0]{\@secondoftwo}%
\providecommand \href [0]{\begingroup \@sanitize@url \@href}%
\providecommand \@href[1]{\@@startlink{#1}\@@href}%
\providecommand \@@href[1]{\endgroup#1\@@endlink}%
\providecommand \@sanitize@url [0]{\catcode `\\12\catcode `\$12\catcode
  `\&12\catcode `\#12\catcode `\^12\catcode `\_12\catcode `\%12\relax}%
\providecommand \@@startlink[1]{}%
\providecommand \@@endlink[0]{}%
\providecommand \url  [0]{\begingroup\@sanitize@url \@url }%
\providecommand \@url [1]{\endgroup\@href {#1}{\urlprefix }}%
\providecommand \urlprefix  [0]{URL }%
\providecommand \Eprint [0]{\href }%
\providecommand \doibase [0]{http://dx.doi.org/}%
\providecommand \selectlanguage [0]{\@gobble}%
\providecommand \bibinfo  [0]{\@secondoftwo}%
\providecommand \bibfield  [0]{\@secondoftwo}%
\providecommand \translation [1]{[#1]}%
\providecommand \BibitemOpen [0]{}%
\providecommand \bibitemStop [0]{}%
\providecommand \bibitemNoStop [0]{.\EOS\space}%
\providecommand \EOS [0]{\spacefactor3000\relax}%
\providecommand \BibitemShut  [1]{\csname bibitem#1\endcsname}%
\let\auto@bib@innerbib\@empty
\bibitem [{\citenamefont {Horsfield}\ \emph {et~al.}(2006)\citenamefont
  {Horsfield}, \citenamefont {Bowler}, \citenamefont {Ness}, \citenamefont
  {S{\'{a}}nchez}, \citenamefont {Todorov},\ and\ \citenamefont
  {Fisher}}]{Horsfield2006}%
  \BibitemOpen
  \bibfield  {author} {\bibinfo {author} {\bibfnamefont {A.~P.}\ \bibnamefont
  {Horsfield}}, \bibinfo {author} {\bibfnamefont {D.~R.}\ \bibnamefont
  {Bowler}}, \bibinfo {author} {\bibfnamefont {H.}~\bibnamefont {Ness}},
  \bibinfo {author} {\bibfnamefont {C.~G.}\ \bibnamefont {S{\'{a}}nchez}},
  \bibinfo {author} {\bibfnamefont {T.~N.}\ \bibnamefont {Todorov}}, \ and\
  \bibinfo {author} {\bibfnamefont {A.~J.}\ \bibnamefont {Fisher}},\ }\href
  {\doibase 10.1088/0034-4885/69/4/R05} {\bibfield  {journal} {\bibinfo
  {journal} {Reports Prog. Phys.}\ }\textbf {\bibinfo {volume} {69}},\ \bibinfo
  {pages} {1195} (\bibinfo {year} {2006})}\BibitemShut {NoStop}%
\bibitem [{\citenamefont {Galperin}\ \emph {et~al.}(2007)\citenamefont
  {Galperin}, \citenamefont {Ratner},\ and\ \citenamefont
  {Nitzan}}]{Galperin2006}%
  \BibitemOpen
  \bibfield  {author} {\bibinfo {author} {\bibfnamefont {M.}~\bibnamefont
  {Galperin}}, \bibinfo {author} {\bibfnamefont {M.~A.}\ \bibnamefont
  {Ratner}}, \ and\ \bibinfo {author} {\bibfnamefont {A.}~\bibnamefont
  {Nitzan}},\ }\href {\doibase 10.1088/0953-8984/19/10/103201} {\bibfield
  {journal} {\bibinfo  {journal} {J. Phys. Condens. Matter}\ }\textbf {\bibinfo
  {volume} {19}},\ \bibinfo {pages} {103201} (\bibinfo {year}
  {2007})}\BibitemShut {NoStop}%
\bibitem [{\citenamefont {Fann}\ \emph
  {et~al.}(1992{\natexlab{a}})\citenamefont {Fann}, \citenamefont {Storz},
  \citenamefont {Tom},\ and\ \citenamefont {Bokor}}]{Fann1992}%
  \BibitemOpen
  \bibfield  {author} {\bibinfo {author} {\bibfnamefont {W.~S.}\ \bibnamefont
  {Fann}}, \bibinfo {author} {\bibfnamefont {R.}~\bibnamefont {Storz}},
  \bibinfo {author} {\bibfnamefont {H.~W.~K.}\ \bibnamefont {Tom}}, \ and\
  \bibinfo {author} {\bibfnamefont {J.}~\bibnamefont {Bokor}},\ }\href
  {\doibase 10.1103/PhysRevLett.68.2834} {\bibfield  {journal} {\bibinfo
  {journal} {Phys. Rev. Lett.}\ }\textbf {\bibinfo {volume} {68}},\ \bibinfo
  {pages} {2834} (\bibinfo {year} {1992}{\natexlab{a}})}\BibitemShut {NoStop}%
\bibitem [{\citenamefont {Fann}\ \emph
  {et~al.}(1992{\natexlab{b}})\citenamefont {Fann}, \citenamefont {Storz},
  \citenamefont {Tom},\ and\ \citenamefont {Bokor}}]{Laboratories1992}%
  \BibitemOpen
  \bibfield  {author} {\bibinfo {author} {\bibfnamefont {W.~S.}\ \bibnamefont
  {Fann}}, \bibinfo {author} {\bibfnamefont {R.}~\bibnamefont {Storz}},
  \bibinfo {author} {\bibfnamefont {H.~W.~K.}\ \bibnamefont {Tom}}, \ and\
  \bibinfo {author} {\bibfnamefont {J.}~\bibnamefont {Bokor}},\ }\href
  {\doibase 10.1103/PhysRevB.46.13592} {\bibfield  {journal} {\bibinfo
  {journal} {Phys. Rev. B}\ }\textbf {\bibinfo {volume} {46}},\ \bibinfo
  {pages} {13592} (\bibinfo {year} {1992}{\natexlab{b}})}\BibitemShut {NoStop}%
\bibitem [{\citenamefont {Ogitsu}\ \emph {et~al.}(2012)\citenamefont {Ogitsu},
  \citenamefont {Ping}, \citenamefont {Correa}, \citenamefont {Cho},
  \citenamefont {Heimann}, \citenamefont {Schwegler}, \citenamefont {Cao},\
  and\ \citenamefont {Collins}}]{Ogitsu2012}%
  \BibitemOpen
  \bibfield  {author} {\bibinfo {author} {\bibfnamefont {T.}~\bibnamefont
  {Ogitsu}}, \bibinfo {author} {\bibfnamefont {Y.}~\bibnamefont {Ping}},
  \bibinfo {author} {\bibfnamefont {A.}~\bibnamefont {Correa}}, \bibinfo
  {author} {\bibfnamefont {B.~I.}\ \bibnamefont {Cho}}, \bibinfo {author}
  {\bibfnamefont {P.}~\bibnamefont {Heimann}}, \bibinfo {author} {\bibfnamefont
  {E.}~\bibnamefont {Schwegler}}, \bibinfo {author} {\bibfnamefont
  {J.}~\bibnamefont {Cao}}, \ and\ \bibinfo {author} {\bibfnamefont {G.~W.}\
  \bibnamefont {Collins}},\ }\href {\doibase 10.1016/j.hedp.2012.01.002}
  {\bibfield  {journal} {\bibinfo  {journal} {High Energy Density Phys.}\
  }\textbf {\bibinfo {volume} {8}},\ \bibinfo {pages} {303} (\bibinfo {year}
  {2012})}\BibitemShut {NoStop}%
\bibitem [{\citenamefont {Duffy}\ \emph {et~al.}(2009)\citenamefont {Duffy},
  \citenamefont {Khakshouri},\ and\ \citenamefont {Rutherford}}]{Duffy2009a}%
  \BibitemOpen
  \bibfield  {author} {\bibinfo {author} {\bibfnamefont {D.}~\bibnamefont
  {Duffy}}, \bibinfo {author} {\bibfnamefont {S.}~\bibnamefont {Khakshouri}}, \
  and\ \bibinfo {author} {\bibfnamefont {A.}~\bibnamefont {Rutherford}},\
  }\href {\doibase 10.1016/j.nimb.2009.06.047} {\bibfield  {journal} {\bibinfo
  {journal} {Nucl. Instruments Methods Phys. Res. Sect. B Beam Interact. with
  Mater. Atoms}\ }\textbf {\bibinfo {volume} {267}},\ \bibinfo {pages} {3050}
  (\bibinfo {year} {2009})}\BibitemShut {NoStop}%
\bibitem [{\citenamefont {H{\"{a}}rtle}\ and\ \citenamefont
  {Thoss}(2011)}]{Hartle2011}%
  \BibitemOpen
  \bibfield  {author} {\bibinfo {author} {\bibfnamefont {R.}~\bibnamefont
  {H{\"{a}}rtle}}\ and\ \bibinfo {author} {\bibfnamefont {M.}~\bibnamefont
  {Thoss}},\ }\href {\doibase 10.1103/PhysRevB.83.125419} {\bibfield  {journal}
  {\bibinfo  {journal} {Phys. Rev. B}\ }\textbf {\bibinfo {volume} {83}},\
  \bibinfo {pages} {125419} (\bibinfo {year} {2011})}\BibitemShut {NoStop}%
\bibitem [{\citenamefont {L{\"{u}}}\ \emph {et~al.}(2012)\citenamefont
  {L{\"{u}}}, \citenamefont {Brandbyge}, \citenamefont {Hedegard},
  \citenamefont {Todorov},\ and\ \citenamefont {Dundas}}]{Lu2012a}%
  \BibitemOpen
  \bibfield  {author} {\bibinfo {author} {\bibfnamefont {J.~T.}\ \bibnamefont
  {L{\"{u}}}}, \bibinfo {author} {\bibfnamefont {M.}~\bibnamefont {Brandbyge}},
  \bibinfo {author} {\bibfnamefont {P.}~\bibnamefont {Hedegard}}, \bibinfo
  {author} {\bibfnamefont {T.~N.}\ \bibnamefont {Todorov}}, \ and\ \bibinfo
  {author} {\bibfnamefont {D.}~\bibnamefont {Dundas}},\ }\href {\doibase
  10.1103/PhysRevB.85.245444} {\bibfield  {journal} {\bibinfo  {journal} {Phys.
  Rev. B}\ }\textbf {\bibinfo {volume} {85}},\ \bibinfo {pages} {245444}
  (\bibinfo {year} {2012})}\BibitemShut {NoStop}%
\bibitem [{\citenamefont {Avigo}\ \emph {et~al.}(2013)\citenamefont {Avigo},
  \citenamefont {Cort{\'{e}}s}, \citenamefont {Rettig}, \citenamefont
  {Thirupathaiah}, \citenamefont {Jeevan}, \citenamefont {Gegenwart},
  \citenamefont {Wolf}, \citenamefont {Ligges}, \citenamefont {Wolf},
  \citenamefont {Fink},\ and\ \citenamefont {Bovensiepen}}]{Avigo2013}%
  \BibitemOpen
  \bibfield  {author} {\bibinfo {author} {\bibfnamefont {I.}~\bibnamefont
  {Avigo}}, \bibinfo {author} {\bibfnamefont {R.}~\bibnamefont {Cort{\'{e}}s}},
  \bibinfo {author} {\bibfnamefont {L.}~\bibnamefont {Rettig}}, \bibinfo
  {author} {\bibfnamefont {S.}~\bibnamefont {Thirupathaiah}}, \bibinfo {author}
  {\bibfnamefont {H.~S.}\ \bibnamefont {Jeevan}}, \bibinfo {author}
  {\bibfnamefont {P.}~\bibnamefont {Gegenwart}}, \bibinfo {author}
  {\bibfnamefont {T.}~\bibnamefont {Wolf}}, \bibinfo {author} {\bibfnamefont
  {M.}~\bibnamefont {Ligges}}, \bibinfo {author} {\bibfnamefont
  {M.}~\bibnamefont {Wolf}}, \bibinfo {author} {\bibfnamefont {J.}~\bibnamefont
  {Fink}}, \ and\ \bibinfo {author} {\bibfnamefont {U.}~\bibnamefont
  {Bovensiepen}},\ }\href {\doibase 10.1088/0953-8984/25/9/094003} {\bibfield
  {journal} {\bibinfo  {journal} {J. Phys. Condens. Matter}\ }\textbf {\bibinfo
  {volume} {25}},\ \bibinfo {pages} {094003} (\bibinfo {year}
  {2013})}\BibitemShut {NoStop}%
\bibitem [{\citenamefont {Lewis}\ \emph {et~al.}(2015)\citenamefont {Lewis},
  \citenamefont {Dong}, \citenamefont {Oliver},\ and\ \citenamefont
  {Fleming}}]{Lewis2015}%
  \BibitemOpen
  \bibfield  {author} {\bibinfo {author} {\bibfnamefont {N.~H.~C.}\
  \bibnamefont {Lewis}}, \bibinfo {author} {\bibfnamefont {H.}~\bibnamefont
  {Dong}}, \bibinfo {author} {\bibfnamefont {T.~A.~A.}\ \bibnamefont {Oliver}},
  \ and\ \bibinfo {author} {\bibfnamefont {G.~R.}\ \bibnamefont {Fleming}},\
  }\href {\doibase 10.1063/1.4919686} {\bibfield  {journal} {\bibinfo
  {journal} {J. Chem. Phys.}\ }\textbf {\bibinfo {volume} {142}},\ \bibinfo
  {pages} {174202} (\bibinfo {year} {2015})}\BibitemShut {NoStop}%
\bibitem [{Note1()}]{Note1}%
  \BibitemOpen
  \bibinfo {note} {Kogoj {\protect \em et al.} arXiv:1509.08431, which appeared
  on arXiv while this paper was under consideration at Physical
  Review.}\BibitemShut {Stop}%
\bibitem [{\citenamefont {Ashcroft}\ and\ \citenamefont
  {Mermin}(1976)}]{Ashcroft}%
  \BibitemOpen
  \bibfield  {author} {\bibinfo {author} {\bibfnamefont {N.~W.}\ \bibnamefont
  {Ashcroft}}\ and\ \bibinfo {author} {\bibfnamefont {D.~N.}\ \bibnamefont
  {Mermin}},\ }\href@noop {} {\emph {\bibinfo {title} {Solid State Physics}}}\
  (\bibinfo  {publisher} {Saunders College},\ \bibinfo {address}
  {Philadelphia},\ \bibinfo {year} {1976})\BibitemShut {NoStop}%
\bibitem [{\citenamefont {Frederiksen}\ \emph {et~al.}(2007)\citenamefont
  {Frederiksen}, \citenamefont {Paulsson}, \citenamefont {Brandbyge},\ and\
  \citenamefont {Jauho}}]{Frederiksen2007}%
  \BibitemOpen
  \bibfield  {author} {\bibinfo {author} {\bibfnamefont {T.}~\bibnamefont
  {Frederiksen}}, \bibinfo {author} {\bibfnamefont {M.}~\bibnamefont
  {Paulsson}}, \bibinfo {author} {\bibfnamefont {M.}~\bibnamefont {Brandbyge}},
  \ and\ \bibinfo {author} {\bibfnamefont {A.-P.}\ \bibnamefont {Jauho}},\
  }\href {\doibase 10.1103/PhysRevB.75.205413} {\bibfield  {journal} {\bibinfo
  {journal} {Phys. Rev. B}\ }\textbf {\bibinfo {volume} {75}},\ \bibinfo
  {pages} {205413} (\bibinfo {year} {2007})}\BibitemShut {NoStop}%
\bibitem [{\citenamefont {Frenkel}\ and\ \citenamefont {Smit}(2002)}]{Frenkel}%
  \BibitemOpen
  \bibfield  {author} {\bibinfo {author} {\bibfnamefont {D.}~\bibnamefont
  {Frenkel}}\ and\ \bibinfo {author} {\bibfnamefont {B.}~\bibnamefont {Smit}},\
  }\href@noop {} {\emph {\bibinfo {title} {Understanding Molecular
  Simulation}}}\ (\bibinfo  {publisher} {Academic Press},\ \bibinfo {year}
  {2002})\BibitemShut {NoStop}%
\bibitem [{\citenamefont {Anisimov}\ \emph {et~al.}(1975)\citenamefont
  {Anisimov}, \citenamefont {Kapeliovich},\ and\ \citenamefont
  {Perel'man}}]{Anisimov1975}%
  \BibitemOpen
  \bibfield  {author} {\bibinfo {author} {\bibfnamefont {S.~I.}\ \bibnamefont
  {Anisimov}}, \bibinfo {author} {\bibfnamefont {B.~L.}\ \bibnamefont
  {Kapeliovich}}, \ and\ \bibinfo {author} {\bibfnamefont {T.~L.}\ \bibnamefont
  {Perel'man}},\ }\href@noop {} {\bibfield  {journal} {\bibinfo  {journal} {J.
  Exp. Theor. Phys.}\ }\textbf {\bibinfo {volume} {39}},\ \bibinfo {pages}
  {375} (\bibinfo {year} {1975})}\BibitemShut {NoStop}%
\bibitem [{\citenamefont {Flynn}\ and\ \citenamefont
  {Averback}(1988)}]{Flynn1988}%
  \BibitemOpen
  \bibfield  {author} {\bibinfo {author} {\bibfnamefont {C.~P.}\ \bibnamefont
  {Flynn}}\ and\ \bibinfo {author} {\bibfnamefont {R.~S.}\ \bibnamefont
  {Averback}},\ }\href {\doibase 10.1103/PhysRevB.38.7118} {\bibfield
  {journal} {\bibinfo  {journal} {Phys. Rev. B}\ }\textbf {\bibinfo {volume}
  {38}},\ \bibinfo {pages} {7118} (\bibinfo {year} {1988})}\BibitemShut
  {NoStop}%
\bibitem [{\citenamefont {Caro}\ and\ \citenamefont
  {Victoria}(1989)}]{Caro1989}%
  \BibitemOpen
  \bibfield  {author} {\bibinfo {author} {\bibfnamefont {A.}~\bibnamefont
  {Caro}}\ and\ \bibinfo {author} {\bibfnamefont {M.}~\bibnamefont
  {Victoria}},\ }\href {\doibase 10.1103/PhysRevA.40.2287} {\bibfield
  {journal} {\bibinfo  {journal} {Phys. Rev. A}\ }\textbf {\bibinfo {volume}
  {40}},\ \bibinfo {pages} {2287} (\bibinfo {year} {1989})}\BibitemShut
  {NoStop}%
\bibitem [{\citenamefont {Finnis}\ \emph {et~al.}(1991)\citenamefont {Finnis},
  \citenamefont {Agnew},\ and\ \citenamefont {Foreman}}]{Finnis1991}%
  \BibitemOpen
  \bibfield  {author} {\bibinfo {author} {\bibfnamefont {M.~W.}\ \bibnamefont
  {Finnis}}, \bibinfo {author} {\bibfnamefont {P.}~\bibnamefont {Agnew}}, \
  and\ \bibinfo {author} {\bibfnamefont {A.~J.~E.}\ \bibnamefont {Foreman}},\
  }\href {\doibase 10.1103/PhysRevB.44.567} {\bibfield  {journal} {\bibinfo
  {journal} {Phys. Rev. B}\ }\textbf {\bibinfo {volume} {44}},\ \bibinfo
  {pages} {567} (\bibinfo {year} {1991})}\BibitemShut {NoStop}%
\bibitem [{\citenamefont {Pr{\"{o}}nnecke}\ \emph {et~al.}(1991)\citenamefont
  {Pr{\"{o}}nnecke}, \citenamefont {Caro}, \citenamefont {Victoria},
  \citenamefont {de~la Rubia},\ and\ \citenamefont {Guinan}}]{Pronnecke1991}%
  \BibitemOpen
  \bibfield  {author} {\bibinfo {author} {\bibfnamefont {S.}~\bibnamefont
  {Pr{\"{o}}nnecke}}, \bibinfo {author} {\bibfnamefont {A.}~\bibnamefont
  {Caro}}, \bibinfo {author} {\bibfnamefont {M.}~\bibnamefont {Victoria}},
  \bibinfo {author} {\bibfnamefont {T.~D.}\ \bibnamefont {de~la Rubia}}, \ and\
  \bibinfo {author} {\bibfnamefont {M.}~\bibnamefont {Guinan}},\ }\href
  {\doibase 10.1557/JMR.1991.0483} {\bibfield  {journal} {\bibinfo  {journal}
  {J. Mater. Res.}\ }\textbf {\bibinfo {volume} {6}},\ \bibinfo {pages} {483}
  (\bibinfo {year} {1991})}\BibitemShut {NoStop}%
\bibitem [{\citenamefont {Duffy}\ and\ \citenamefont
  {Rutherford}(2007)}]{Duffy2007}%
  \BibitemOpen
  \bibfield  {author} {\bibinfo {author} {\bibfnamefont {D.~M.}\ \bibnamefont
  {Duffy}}\ and\ \bibinfo {author} {\bibfnamefont {A.~M.}\ \bibnamefont
  {Rutherford}},\ }\href {\doibase 10.1088/0953-8984/19/1/016207} {\bibfield
  {journal} {\bibinfo  {journal} {J. Phys. Condens. Matter}\ }\textbf {\bibinfo
  {volume} {19}},\ \bibinfo {pages} {016207} (\bibinfo {year}
  {2007})}\BibitemShut {NoStop}%
\bibitem [{\citenamefont {Race}\ \emph {et~al.}(2010)\citenamefont {Race},
  \citenamefont {Mason}, \citenamefont {Finnis}, \citenamefont {Foulkes},
  \citenamefont {Horsfield},\ and\ \citenamefont {Sutton}}]{Race2010}%
  \BibitemOpen
  \bibfield  {author} {\bibinfo {author} {\bibfnamefont {C.~P.}\ \bibnamefont
  {Race}}, \bibinfo {author} {\bibfnamefont {D.~R.}\ \bibnamefont {Mason}},
  \bibinfo {author} {\bibfnamefont {M.~W.}\ \bibnamefont {Finnis}}, \bibinfo
  {author} {\bibfnamefont {W.~M.~C.}\ \bibnamefont {Foulkes}}, \bibinfo
  {author} {\bibfnamefont {A.~P.}\ \bibnamefont {Horsfield}}, \ and\ \bibinfo
  {author} {\bibfnamefont {A.~P.}\ \bibnamefont {Sutton}},\ }\href {\doibase
  10.1088/0034-4885/73/11/116501} {\bibfield  {journal} {\bibinfo  {journal}
  {Reports Prog. Phys.}\ }\textbf {\bibinfo {volume} {73}},\ \bibinfo {pages}
  {116501} (\bibinfo {year} {2010})}\BibitemShut {NoStop}%
\bibitem [{\citenamefont {Mason}(2015)}]{Mason2015}%
  \BibitemOpen
  \bibfield  {author} {\bibinfo {author} {\bibfnamefont {D.}~\bibnamefont
  {Mason}},\ }\href {\doibase 10.1088/0953-8984/27/14/145401} {\bibfield
  {journal} {\bibinfo  {journal} {J. Phys. Condens. Matter}\ }\textbf {\bibinfo
  {volume} {27}},\ \bibinfo {pages} {145401} (\bibinfo {year}
  {2015})}\BibitemShut {NoStop}%
\bibitem [{\citenamefont {Cho}\ \emph {et~al.}(2011)\citenamefont {Cho},
  \citenamefont {Engelhorn}, \citenamefont {Correa}, \citenamefont {Ogitsu},
  \citenamefont {Weber}, \citenamefont {Lee}, \citenamefont {Feng},
  \citenamefont {Ni}, \citenamefont {Ping}, \citenamefont {Nelson},
  \citenamefont {Prendergast}, \citenamefont {Lee}, \citenamefont {Falcone},\
  and\ \citenamefont {Heimann}}]{Cho2011}%
  \BibitemOpen
  \bibfield  {author} {\bibinfo {author} {\bibfnamefont {B.~I.}\ \bibnamefont
  {Cho}}, \bibinfo {author} {\bibfnamefont {K.}~\bibnamefont {Engelhorn}},
  \bibinfo {author} {\bibfnamefont {A.~A.}\ \bibnamefont {Correa}}, \bibinfo
  {author} {\bibfnamefont {T.}~\bibnamefont {Ogitsu}}, \bibinfo {author}
  {\bibfnamefont {C.~P.}\ \bibnamefont {Weber}}, \bibinfo {author}
  {\bibfnamefont {H.~J.}\ \bibnamefont {Lee}}, \bibinfo {author} {\bibfnamefont
  {J.}~\bibnamefont {Feng}}, \bibinfo {author} {\bibfnamefont {P.~A.}\
  \bibnamefont {Ni}}, \bibinfo {author} {\bibfnamefont {Y.}~\bibnamefont
  {Ping}}, \bibinfo {author} {\bibfnamefont {A.~J.}\ \bibnamefont {Nelson}},
  \bibinfo {author} {\bibfnamefont {D.}~\bibnamefont {Prendergast}}, \bibinfo
  {author} {\bibfnamefont {R.~W.}\ \bibnamefont {Lee}}, \bibinfo {author}
  {\bibfnamefont {R.~W.}\ \bibnamefont {Falcone}}, \ and\ \bibinfo {author}
  {\bibfnamefont {P.~A.}\ \bibnamefont {Heimann}},\ }\href {\doibase
  10.1103/PhysRevLett.106.167601} {\bibfield  {journal} {\bibinfo  {journal}
  {Phys. Rev. Lett.}\ }\textbf {\bibinfo {volume} {106}},\ \bibinfo {pages}
  {167601} (\bibinfo {year} {2011})}\BibitemShut {NoStop}%
\bibitem [{\citenamefont {Karim}\ \emph {et~al.}(2014)\citenamefont {Karim},
  \citenamefont {Shugaev}, \citenamefont {Wu}, \citenamefont {Lin},
  \citenamefont {Hainsey},\ and\ \citenamefont {Zhigilei}}]{Karim2014}%
  \BibitemOpen
  \bibfield  {author} {\bibinfo {author} {\bibfnamefont {E.~T.}\ \bibnamefont
  {Karim}}, \bibinfo {author} {\bibfnamefont {M.}~\bibnamefont {Shugaev}},
  \bibinfo {author} {\bibfnamefont {C.}~\bibnamefont {Wu}}, \bibinfo {author}
  {\bibfnamefont {Z.}~\bibnamefont {Lin}}, \bibinfo {author} {\bibfnamefont
  {R.~F.}\ \bibnamefont {Hainsey}}, \ and\ \bibinfo {author} {\bibfnamefont
  {L.~V.}\ \bibnamefont {Zhigilei}},\ }\href@noop {} {\bibfield  {journal}
  {\bibinfo  {journal} {J. Appl. Phys.}\ }\textbf {\bibinfo {volume} {115}}
  (\bibinfo {year} {2014})}\BibitemShut {NoStop}%
\bibitem [{\citenamefont {Zarkadoula}\ \emph {et~al.}(2014)\citenamefont
  {Zarkadoula}, \citenamefont {Daraszewicz}, \citenamefont {Duffy},
  \citenamefont {Seaton}, \citenamefont {Todorov}, \citenamefont {Nordlund},
  \citenamefont {Dove},\ and\ \citenamefont {Trachenko}}]{Zarkadoula2014a}%
  \BibitemOpen
  \bibfield  {author} {\bibinfo {author} {\bibfnamefont {E.}~\bibnamefont
  {Zarkadoula}}, \bibinfo {author} {\bibfnamefont {S.~L.}\ \bibnamefont
  {Daraszewicz}}, \bibinfo {author} {\bibfnamefont {D.~M.}\ \bibnamefont
  {Duffy}}, \bibinfo {author} {\bibfnamefont {M.~A.}\ \bibnamefont {Seaton}},
  \bibinfo {author} {\bibfnamefont {I.~T.}\ \bibnamefont {Todorov}}, \bibinfo
  {author} {\bibfnamefont {K.}~\bibnamefont {Nordlund}}, \bibinfo {author}
  {\bibfnamefont {M.~T.}\ \bibnamefont {Dove}}, \ and\ \bibinfo {author}
  {\bibfnamefont {K.}~\bibnamefont {Trachenko}},\ }\href {\doibase
  10.1088/0953-8984/26/8/085401} {\bibfield  {journal} {\bibinfo  {journal} {J.
  Phys. Condens. Matter}\ }\textbf {\bibinfo {volume} {26}},\ \bibinfo {pages}
  {085401} (\bibinfo {year} {2014})}\BibitemShut {NoStop}%
\bibitem [{\citenamefont {Horsfield}\ \emph {et~al.}(2004)\citenamefont
  {Horsfield}, \citenamefont {Bowler}, \citenamefont {Fisher}, \citenamefont
  {Todorov},\ and\ \citenamefont {S{\'{a}}nchez}}]{Horsfield2004}%
  \BibitemOpen
  \bibfield  {author} {\bibinfo {author} {\bibfnamefont {A.~P.}\ \bibnamefont
  {Horsfield}}, \bibinfo {author} {\bibfnamefont {D.~R.}\ \bibnamefont
  {Bowler}}, \bibinfo {author} {\bibfnamefont {A.~J.}\ \bibnamefont {Fisher}},
  \bibinfo {author} {\bibfnamefont {T.~N.}\ \bibnamefont {Todorov}}, \ and\
  \bibinfo {author} {\bibfnamefont {C.~G.}\ \bibnamefont {S{\'{a}}nchez}},\
  }\href {\doibase 10.1088/0953-8984/16/46/012} {\bibfield  {journal} {\bibinfo
   {journal} {J. Phys. Condens. Matter}\ }\textbf {\bibinfo {volume} {16}},\
  \bibinfo {pages} {8251} (\bibinfo {year} {2004})}\BibitemShut {NoStop}%
\bibitem [{\citenamefont {Horsfield}\ \emph {et~al.}(2005)\citenamefont
  {Horsfield}, \citenamefont {Bowler}, \citenamefont {Fisher}, \citenamefont
  {Todorov},\ and\ \citenamefont {S{\'{a}}nchez}}]{Horsfield2005}%
  \BibitemOpen
  \bibfield  {author} {\bibinfo {author} {\bibfnamefont {A.~P.}\ \bibnamefont
  {Horsfield}}, \bibinfo {author} {\bibfnamefont {D.~R.}\ \bibnamefont
  {Bowler}}, \bibinfo {author} {\bibfnamefont {A.~J.}\ \bibnamefont {Fisher}},
  \bibinfo {author} {\bibfnamefont {T.~N.}\ \bibnamefont {Todorov}}, \ and\
  \bibinfo {author} {\bibfnamefont {C.~G.}\ \bibnamefont {S{\'{a}}nchez}},\
  }\href {\doibase 10.1088/0953-8984/17/30/006} {\bibfield  {journal} {\bibinfo
   {journal} {J. Phys. Condens. Matter}\ }\textbf {\bibinfo {volume} {17}},\
  \bibinfo {pages} {4793} (\bibinfo {year} {2005})}\BibitemShut {NoStop}%
\bibitem [{\citenamefont {Stella}\ \emph {et~al.}(2007)\citenamefont {Stella},
  \citenamefont {Meister}, \citenamefont {Fisher},\ and\ \citenamefont
  {Horsfield}}]{Stella2007}%
  \BibitemOpen
  \bibfield  {author} {\bibinfo {author} {\bibfnamefont {L.}~\bibnamefont
  {Stella}}, \bibinfo {author} {\bibfnamefont {M.}~\bibnamefont {Meister}},
  \bibinfo {author} {\bibfnamefont {A.~J.}\ \bibnamefont {Fisher}}, \ and\
  \bibinfo {author} {\bibfnamefont {A.~P.}\ \bibnamefont {Horsfield}},\ }\href
  {\doibase 10.1063/1.2801537} {\bibfield  {journal} {\bibinfo  {journal} {J.
  Chem. Phys.}\ }\textbf {\bibinfo {volume} {127}},\ \bibinfo {pages} {214104}
  (\bibinfo {year} {2007})}\BibitemShut {NoStop}%
\bibitem [{\citenamefont {Wang}\ \emph {et~al.}(2015)\citenamefont {Wang},
  \citenamefont {Long},\ and\ \citenamefont {Prezhdo}}]{Wang2015}%
  \BibitemOpen
  \bibfield  {author} {\bibinfo {author} {\bibfnamefont {L.}~\bibnamefont
  {Wang}}, \bibinfo {author} {\bibfnamefont {R.}~\bibnamefont {Long}}, \ and\
  \bibinfo {author} {\bibfnamefont {O.~V.}\ \bibnamefont {Prezhdo}},\ }\href
  {\doibase 10.1146/annurev-physchem-040214-121359} {\bibfield  {journal}
  {\bibinfo  {journal} {Annu. Rev. Phys. Chem.}\ }\textbf {\bibinfo {volume}
  {66}},\ \bibinfo {pages} {549} (\bibinfo {year} {2015})}\BibitemShut
  {NoStop}%
\bibitem [{\citenamefont {Cahill}\ \emph {et~al.}(2014)\citenamefont {Cahill},
  \citenamefont {Braun}, \citenamefont {Chen}, \citenamefont {Clarke},
  \citenamefont {Fan}, \citenamefont {Goodson}, \citenamefont {Keblinski},
  \citenamefont {King}, \citenamefont {Mahan}, \citenamefont {Majumdar},
  \citenamefont {Maris}, \citenamefont {Phillpot}, \citenamefont {Pop},\ and\
  \citenamefont {Shi}}]{Cahill2014a}%
  \BibitemOpen
  \bibfield  {author} {\bibinfo {author} {\bibfnamefont {D.~G.}\ \bibnamefont
  {Cahill}}, \bibinfo {author} {\bibfnamefont {P.~V.}\ \bibnamefont {Braun}},
  \bibinfo {author} {\bibfnamefont {G.}~\bibnamefont {Chen}}, \bibinfo {author}
  {\bibfnamefont {D.~R.}\ \bibnamefont {Clarke}}, \bibinfo {author}
  {\bibfnamefont {S.}~\bibnamefont {Fan}}, \bibinfo {author} {\bibfnamefont
  {K.~E.}\ \bibnamefont {Goodson}}, \bibinfo {author} {\bibfnamefont
  {P.}~\bibnamefont {Keblinski}}, \bibinfo {author} {\bibfnamefont {W.~P.}\
  \bibnamefont {King}}, \bibinfo {author} {\bibfnamefont {G.~D.}\ \bibnamefont
  {Mahan}}, \bibinfo {author} {\bibfnamefont {A.}~\bibnamefont {Majumdar}},
  \bibinfo {author} {\bibfnamefont {H.~J.}\ \bibnamefont {Maris}}, \bibinfo
  {author} {\bibfnamefont {S.~R.}\ \bibnamefont {Phillpot}}, \bibinfo {author}
  {\bibfnamefont {E.}~\bibnamefont {Pop}}, \ and\ \bibinfo {author}
  {\bibfnamefont {L.}~\bibnamefont {Shi}},\ }\href
  {http://scitation.aip.org/content/aip/journal/apr2/1/1/10.1063/1.4832615}
  {\bibfield  {journal} {\bibinfo  {journal} {Appl. Phys. Rev.}\ }\textbf
  {\bibinfo {volume} {1}},\ \bibinfo {pages} {011305} (\bibinfo {year}
  {2014})}\BibitemShut {NoStop}%
\bibitem [{\citenamefont {Mozyrsky}\ and\ \citenamefont
  {Martin}(2002)}]{Mozyrsky2002}%
  \BibitemOpen
  \bibfield  {author} {\bibinfo {author} {\bibfnamefont {D.}~\bibnamefont
  {Mozyrsky}}\ and\ \bibinfo {author} {\bibfnamefont {I.}~\bibnamefont
  {Martin}},\ }\href {\doibase 10.1103/PhysRevLett.89.018301} {\bibfield
  {journal} {\bibinfo  {journal} {Phys. Rev. Lett.}\ }\textbf {\bibinfo
  {volume} {89}},\ \bibinfo {pages} {018301} (\bibinfo {year}
  {2002})}\BibitemShut {NoStop}%
\bibitem [{\citenamefont {Todorov}\ \emph {et~al.}(2014)\citenamefont
  {Todorov}, \citenamefont {Dundas}, \citenamefont {L{\"{u}}}, \citenamefont
  {Brandbyge},\ and\ \citenamefont {Hedegard}}]{Todorov2014}%
  \BibitemOpen
  \bibfield  {author} {\bibinfo {author} {\bibfnamefont {T.~N.}\ \bibnamefont
  {Todorov}}, \bibinfo {author} {\bibfnamefont {D.}~\bibnamefont {Dundas}},
  \bibinfo {author} {\bibfnamefont {J.-T.}\ \bibnamefont {L{\"{u}}}}, \bibinfo
  {author} {\bibfnamefont {M.}~\bibnamefont {Brandbyge}}, \ and\ \bibinfo
  {author} {\bibfnamefont {P.}~\bibnamefont {Hedegard}},\ }\href {\doibase
  10.1088/0143-0807/35/6/065004} {\bibfield  {journal} {\bibinfo  {journal}
  {Eur. J. Phys.}\ }\textbf {\bibinfo {volume} {35}},\ \bibinfo {pages}
  {065004} (\bibinfo {year} {2014})}\BibitemShut {NoStop}%
\bibitem [{Note2()}]{Note2}%
  \BibitemOpen
  \bibinfo {note} {Here we ignore the additional term $\protect \mathaccentV
  {hat}05E{\rho }_\protect \mathrm {e}(t)\protect \tmspace +\thinmuskip
  {.1667em}\protect \mathrm {Tr}_{\protect \mathrm {e}} ( \protect \mathaccentV
  {hat}05E{F} \protect \mathaccentV {hat}05E{\rho }_\protect \mathrm {e}(t) )$.
  It corresponds to the so-called ``Hartree'' diagram in NEGF treatments of
  electron-phonon interactions \cite {Frederiksen2007}, and is related to
  motion of the oscillator centroid, a mean-field property. This term involves
  the mean force $\protect \mathrm {Tr}_{\protect \mathrm {e}} ( \protect
  \mathaccentV {hat}05E{F} \protect \mathaccentV {hat}05E{\rho }_\protect
  \mathrm {e}(t) )$ on a given degree of freedom, which in the present examples
  is orders of magnitude less than a typical interatomic bond
  force.}\BibitemShut {Stop}%
\bibitem [{\citenamefont {Burke}\ \emph {et~al.}(2005)\citenamefont {Burke},
  \citenamefont {Car},\ and\ \citenamefont {Gebauer}}]{Burke2005}%
  \BibitemOpen
  \bibfield  {author} {\bibinfo {author} {\bibfnamefont {K.}~\bibnamefont
  {Burke}}, \bibinfo {author} {\bibfnamefont {R.}~\bibnamefont {Car}}, \ and\
  \bibinfo {author} {\bibfnamefont {R.}~\bibnamefont {Gebauer}},\ }\href
  {\doibase 10.1103/PhysRevLett.94.146803} {\bibfield  {journal} {\bibinfo
  {journal} {Phys. Rev. Lett.}\ }\textbf {\bibinfo {volume} {94}},\ \bibinfo
  {pages} {146803} (\bibinfo {year} {2005})}\BibitemShut {NoStop}%
\bibitem [{\citenamefont {L{\"{u}}}\ \emph {et~al.}(2011)\citenamefont
  {L{\"{u}}}, \citenamefont {Hedegard},\ and\ \citenamefont
  {Brandbyge}}]{Lu2011}%
  \BibitemOpen
  \bibfield  {author} {\bibinfo {author} {\bibfnamefont {J.~T.}\ \bibnamefont
  {L{\"{u}}}}, \bibinfo {author} {\bibfnamefont {P.}~\bibnamefont {Hedegard}},
  \ and\ \bibinfo {author} {\bibfnamefont {M.}~\bibnamefont {Brandbyge}},\
  }\href {\doibase 10.1103/PhysRevLett.107.046801} {\bibfield  {journal}
  {\bibinfo  {journal} {Phys. Rev. Lett.}\ }\textbf {\bibinfo {volume} {107}},\
  \bibinfo {pages} {046801} (\bibinfo {year} {2011})}\BibitemShut {NoStop}%
\bibitem [{\citenamefont {Theilhaber}(1992)}]{Theilhaber1992}%
  \BibitemOpen
  \bibfield  {author} {\bibinfo {author} {\bibfnamefont {J.}~\bibnamefont
  {Theilhaber}},\ }\href {\doibase 10.1103/PhysRevB.46.12990} {\bibfield
  {journal} {\bibinfo  {journal} {Phys. Rev. B}\ }\textbf {\bibinfo {volume}
  {46}},\ \bibinfo {pages} {12990} (\bibinfo {year} {1992})}\BibitemShut
  {NoStop}%
\end{thebibliography}%

\end{document}